\documentclass{llncs}

\usepackage{natbib}
\usepackage{graphicx}

\usepackage{fancyhdr}
\usepackage{lastpage}
 
\begin{document}

\title{A short review and primer on eye tracking in human computer interaction applications}
\author{Kristian Lukander}
\institute{Quantified Employee unit, Finnish Institute of Occupational Health,\\
\email{kristian.lukander@ttl.fi},\\
POBox 40, Helsinki, 00250, Finland}

\maketitle              % typeset the title of the contribution

\begin{abstract}
The application of psychophysiologicy in human-computer interaction is a growing field with significant potential for future smart personalised systems. Working in this emerging field requires comprehension of an array of physiological signals and analysis techniques. 

Eye tracking is a widely used method for tracking user attention with gaze location, but also provides information on the internal cognitive and contextual state, intention, and the locus of the user's visual attention in interactive settings through a number of eye and eyelid movement related parameters. This paper presents a short review on the application of eye tracking in human-computer interaction.

This paper aims to serve as a primer for the novice, enabling rapid familiarisation with the latest core concepts. We put special emphasis on everyday human-computer interface applications to distinguish from the more common clinical or sports uses of psychophysiology.

This paper is an extract from a comprehensive review of the entire field of ambulatory psychophysiology, including 12 similar chapters, plus application guidelines and systematic review. Thus any citation should be made using the following reference:

{\parshape 1 2cm \dimexpr\linewidth-1cm\relax
B. Cowley, M. Filetti, K. Lukander, J. Torniainen, A. Henelius, L. Ahonen, O. Barral, I. Kosunen, T. Valtonen, M. Huotilainen, N. Ravaja, G. Jacucci. \textit{The Psychophysiology Primer: a guide to methods and a broad review with a focus on human-computer interaction}. Foundations and Trends in Human-Computer Interaction, vol. 9, no. 3-4, pp. 150--307, 2016.
\par}


\end{abstract}

As the proverbial `windows of the soul', the eyes can be considered the only surface feature of the central nervous system. As such, they should be a rich source of information on human cognition and activity. In this brief review, we consider eye/gaze tracking as a source of information on internal cognitive and contextual state, intention, and the locus of the user's visual attention in interactive settings. Because the primary function of the eye is to sample visual information from the environment, re-purposing it for direct interaction, by using gaze as an `input method', has proved to be problematic and results in a phenomenon known as the Midas touch \citep{jacob1990}: attentive information-seeking fixations are interpreted too readily as actions within the gaze-controlled interface. Eye tracking in direct-interaction applications has been discussed recently by \citet{fairclough2014} and is not discussed further here.

\section{Background}

Eye movements are highly task- and context-specific \citep{rothkopf2015}. In a task performance context, eye movements can be divided into fixational and saccadic movements. Fixational control (consisting of fixations and smooth pursuit movements that track slowly moving targets) is aimed at stabilising the retinal image for information extraction, whereas saccadic movements rapidly move the point of gaze to fixate and acquire new features from within the visual field. Vergence (convergent, independent movement of the eyes) can provide additional information on the depth plane of visual focus in binocular viewing. 

In a fairly novel development, also micro-saccades, minor directed movements of the eye within a fixation, have been shown to correlate with cognitive activity, especially visuo-spatial attention \citep{meyberg2015}. However, as tracking these requires special equipment that is not usable in everyday contexts, our review will deal only with the eye movements listed above.

Eye trackers typically enable the extraction of several parameters for each type of eye movement. For fixations, typical parameters are location, duration, frequency, and drift within fixations. For saccades, the usual parameters considered are frequency, duration, amplitude, average speed, and speed profiles. In addition to eye movements, eye tracking can provide information on blinks, and typical parameters in this connection include frequency, closing of the eye(s), and eyelid closing and opening times. Among the more complex, derived parameters are dwell times (the sum of fixation times within a defined area of interest); gaze paths and patterns; the area covered; and the frequency, number of, and sequence of areas of interest visited in visual stimuli or an interface. 

\section{Methods}

The two most firmly established eye tracking techniques at present are the electro-ogulogram and video-oculography. The first of these is generated from measurement of electrical activity associated with eye movements, quantified by recording from electrodes applied to the skin surface around the eye(s). The EOG approach provides high temporal resolution (up to several kilohertz) and can even be used when the eyes are closed (during sleep or at sleep onset), though it offers limited spatial resolution, has a drifting baseline, and exhibits high-frequency noise \citep{eggert2007}. These properties make EOGs suitable for wearable devices tracking oculomotor parameters, but they are less useful in actual point-of-gaze (POG) tracking. Although EOG measurement set-ups vary, most often the measurements are performed with four electrodes: two horizontal, placed at the outer canthi of the eyes, with the signal summing the movement of the electrical dipoles of the two eyes, and two vertical, placed above and below one eye, which can be used also for tracking blinks. 

Devices for VOG measurements are camera-based, tracking the movements of the eye via changes in visual features such as the pupil, iris, sclera, eyelid, and light source reflections on the surface of the cornea. Pupillary measures available through VOG have been linked with cognitive activity, and these are discussed in \citep[sect.3.9]{Cowley2016primer}. Video-oculography devices, which can be grouped into wearable and desktop devices, vary significantly in their capabilities and requirements. Desktop devices encompass both very precise trackers, requiring the use of head and chin rests, and completely non-intrusive remote trackers that may be integrated into laptop computers and monitors. While VOG devices provide better spatial resolution than does EOG (typically 0.5--2 degrees of visual angle), all but the most expensive tracking systems offer only limited temporal resolution (30--60~Hz). Hence, VOG is better suited to tracking the POG, examining gaze patterns, and utilising event-based metrics. 

\begin{figure}[t]
\begin{center}
\includegraphics{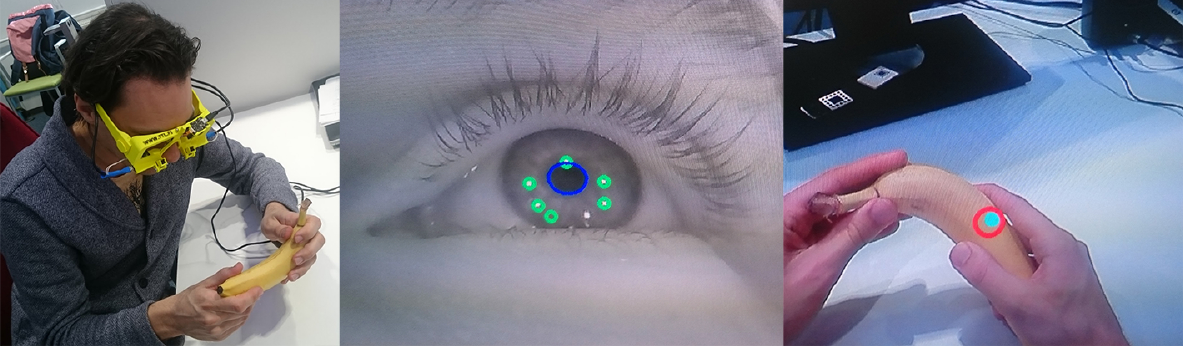}
\end{center}
\caption{A mobile gaze tracking device in an ecologically valid application, in the left panel. The centre panel provides a visualisation of the algorithm's performance: the smaller (green) ellipses show detection of infrared light source reflections; the larger (blue) ellipse shows the subsequently detected pupil. The right panel shows the consequent detection of the point of gaze. See \citet{lukander2013} for details.}
\label{fig.eyetracking}
\end{figure}

\section{Applications}
We will now present several example applications of eye tracking for establishing the internal state of the user on the basis of eye and gaze metrics. 

\subsection{Mental workload}
Mental workload has been studied under various task conditions. \citet{hutton2008} presented examples of saccadic eye movements being affected by working memory and attention requirements in experimental stimulus--response paradigms. \citet{chang2014} used gaze duration (dwell times on sentences) for measuring the spatial and temporal distribution of attention in a reading task, with `seductive details' grabbing attention and leading to poor comprehension and recall performance.

\citet{tokuda2011} introduced the detection of `saccadic intrusions' -- quick fixation-restoring saccades -- and their frequency for estimating mental workload, while \citet{bodala2014} utilised peak saccade velocity for gauging mental workload, claiming that higher cognitive workload necessitates faster saccades. 

\subsection{Flow and focus}
Successfully tracking flow and focus, and thereby being able to support users' task engagement, could offer a highly productive use for eye tracking. \citet{tsai2007} reported that in a dual visual-auditory task condition, increased fixation frequency was indicative of a focus on the visual task while reduced blink frequency and horizontal vergence indicated a focus on the auditory task. \citet{wang2014} explored gaze patterns during driving, concluding that gaze patterns contract in conditions of increased cognitive demands, focusing on a more limited area around the item of attention. \citet{dehais2012}, in the meantime, suggested an `index of attentional focus', based on decreased saccadic activity and more concentrated fixations. 

\citet{marshall2007} described seven binary eye-based metrics, which she used for studying cognitive state changes in three set-ups: relaxed/engaged problem-solving, focused/distracted attention in driving, and alert/fatigued state during performance of a visual search. Marshall reported that all seven eye metrics successfully discriminated among the states, with classification accuracies between 67\% and 93\%. 

\subsection{Fatigue/sleepiness}
Fatigue and sleepiness studies have concentrated mainly on EOG, for obvious practical reasons -- both blink and saccade parameters derived from EOG have been shown to be sensitive to fatigue caused by sleepiness.
For blinks, \citet{barbato1995, khushaba2011} used blink rates and \citet{ingre2006, papadelis2007} used blink duration; while \citet{morris1996} showed that blink amplitude and eye closing times change significantly in consequence of fatigue. For saccades, \citet{wright2001, hyoki1998} studied saccade rate and eye activity metrics, and \citet{morris1996, hyoki1998, thomas2007} used saccadic velocity parameters for detecting fatigue and sleepiness. Another study suggested that the peak velocity of horizontal saccades could be the most powerful eye parameter for monitoring sleepiness \citep{hirvonen2010}. \citet{thomas2007} presented saccadic velocity as an oculomotor neurophysiological measure significantly correlated with decreasing brain metabolism and cognitive performance, thereby demonstrating that it could be used as a surrogate marker for the for the cognitive state of alertness or sleep deprivation.

\subsection{Application-specificity}
A large proportion of eye movement research to date has been performed in limited laboratory environments. One of the pioneers of eye tracking research, \citet{rayner2009}, warns also that it might be hazardous to generalise eye movement metrics across task types such as reading and visual search. As eye movement metrics are highly task- and subject-specific, movements in the real world can ultimately be understood only in the context of a particular task. 

\section{Conclusions}
Metrics of eye movements provide a rich, contextual source of information on human behaviour and internal cognitive states, applicable to various HCI endeavours. Recent developments in measuring and analysing ocular behaviour can inform novel tools, enabling their use in natural everyday environments. As a large proportion of the existing studies have looked at eye tracking in laboratory conditions, studying and applying gaze interaction and gaze-based user modelling in natural environments presents a substantial opportunity. However, individual-to-individual variability and the task-specific nature of eye movements should be carefully considered, if one is to deliver successful applications of eye-aware user interfaces and insights into the cognitive state of users. 

\bibliographystyle{plainnat}
\bibliography{preprint_eyetracking}

\end{document}